\documentclass[runningheads]{llncs}
\usepackage{amsmath}
\usepackage{graphicx}
\usepackage{hyperref}
\usepackage{courier}
\usepackage{multicol}
\usepackage{hyperref}
\usepackage{academicons}
\usepackage{xcolor}
\newcommand{\orcid}[1]{\href{https://orcid.org/#1}{\textcolor[HTML]{A6CE39}{\aiOrcid}}}

\begin{document}
\title{AutoPET Challenge: Combining nn-Unet with Swin UNETR Augmented by Maximum Intensity Projection Classifier}
%
\titlerunning{AutoPET: Combining nn-Unet, Swin UNETR, and MIP Classifiers}
%
\author{Lars Heiliger\inst{1,2}$^{,*}$ \and
Zdravko Marinov\inst{3,9}$^{,*}$ \and
Max Hasin \inst{1} \and
André Ferreira\inst{1,4,5} \and
Jana Fragemann\inst{1,2} \and
Kelsey Pomykala\inst{1} \and
Jacob Murray\inst{1} \and
David Kersting\inst{6,7,8} \and
Victor Alves\inst{4}\and
Rainer Stiefelhagen\inst{3} \and
Jan Egger\inst{1,2}$^{,**}$ \and
Jens Kleesiek\inst{1,2}$^{,**}$ }
\authorrunning{L. Heiliger et al.}
%
\institute{Institute for AI in Medicine, University Hospital Essen, Essen, Germany \email{\{lars.heiliger, jan.egger, jens.kleesiek\}@uk-essen.de} \and
Cancer Research Center Cologne Essen (CCCE), University Medicine Essen, Essen, Germany \and
Institute for Anthropomatics and Robotics, Karlsruhe Institute of Technology, Karlsruhe, Germany  \and
Center Algoritmi, University of Minho, Braga, Portugal \and Computer Algorithms for Medicine Laboratory, Graz, Austria \and Department of Nuclear Medicine, University Hospital Essen, Essen, Germany \and German Cancer Consortium (DKTK), Essen, Germany \and West German Cancer Center, Essen, Germany \and HIDSS4Health - Helmholtz Information and Data Science School for Health, Karlsruhe/Heidelberg, Germany
}

\def\thefootnote{*}\footnotetext{These authors contributed equally to this work}\def\thefootnote{\arabic{footnote}}

\def\thefootnote{**}\footnotetext{Shared last author}\def\thefootnote{\arabic{footnote}}

\maketitle              
\begin{abstract}
Tumor volume and changes in tumor characteristics over time are important biomarkers for cancer therapy. In this context, FDG-PET/CT scans are routinely used for staging and re-staging of cancer, as the radiolabeled fluorodeoxyglucose is taken up in regions of high metabolism. Unfortunately, these regions with high metabolism are not specific to tumors and can also represent physiological uptake by normal functioning organs, inflammation, or infection, making detailed and reliable tumor segmentation in these scans a demanding task. This gap in research is addressed by the AutoPET challenge, which provides a public data set with FDG-PET/CT scans from 900 patients to encourage further improvement in this field.
Our contribution to this challenge is an ensemble of two state-of-the-art segmentation models, the nn-Unet and the Swin UNETR,  augmented by a maximum intensity projection classifier that acts like a gating mechanism. If it predicts the existence of lesions, both segmentations are combined by a late fusion approach. Our solution achieves an average Cross Validation Dice score of 72.12\% on patients diagnosed with lung cancer, melanoma, and lymphoma in our cross-validation.\\
Code: \url{https://github.com/heiligerl/AutoPET_Challenge_Submission}
\end{abstract}
\section{Introduction}
\label{sec:introduction}
Medical imaging is crucial to detect and assess the progress of cancer and medical deep learning can help to automate this time-consuming assessment \cite{egger2022medical}. In the clinical routine, positron emission tomography / computed tomography (PET/CT) enables the visualization of metabolic processes inside tissues. In the oncologic setting, fluorodeoxyglucose (FDG) is the most widely used PET tracer, which can display the glucose consumption of tissues. Considering malignant solid tumor entities, lesions typically exhibit an increased glucose consumption.

Due to the lack of automated solutions for the analysis of FDG-PET/CT imaging, which would possibly allow for more precise and individualized treatment decisions, experienced physicians analyze the images in a qualitative way.

The AutoPET challenge\footnote{\url{https://autopet.grand-challenge.org/}} \cite{sergios_gatidis_2022_6362493} addresses this gap by providing a large and publicly available training data set \cite{dataset} with the aim of enhancing the development of 3D semantic segmentation models subject to the avoidance of false positive segmentations.

The training dataset consists of 1,014 PET/CT scans from 900 patients from the University Hospital Tübingen, Germany. There are 513 scans without lesions, and 188, 168, and 145 scans are associated with melanoma, lung cancer, and lymphoma, respectively.

The challenge's test set consists of 200 PET/CT scans collected from the University Hospital Tübingen (100 scans) and the University Hospital of Ludwig-Maximilians-Universität (LMU) Munich, Germany, (100 scans). The test set is hidden from the competitors, and the containerized models have to be submitted  to the Grand Challenge platform\footnote{\url{https://grand-challenge.org/}}.
The evaluation procedure applies three different metrics: (1) the foreground Dice score of segmented lesions, (2) the volume of false positive connected components that do not overlap with positives (=false positive volume, FPV), and (3) the volume of positive connected components in the ground truth that do not overlap with the estimated segmentation (=false negative volume, FNV).

  When evaluating developed models on the hidden test set, all three metrics are considered for non-healthy patients, whereas only FPV is considered for healthy cases.
The final leader board position is based on the ranks in these three metrics with the weighting (0.5, 0.25, 0.25).

We present the methodology of our solution in the subsequent section.
\section{Methodology}
\subsection{Pipeline of Our Final Submission}\label{sec:pipeline}
\begin{figure}[!h]
    \centering
    \includegraphics[scale=0.45]{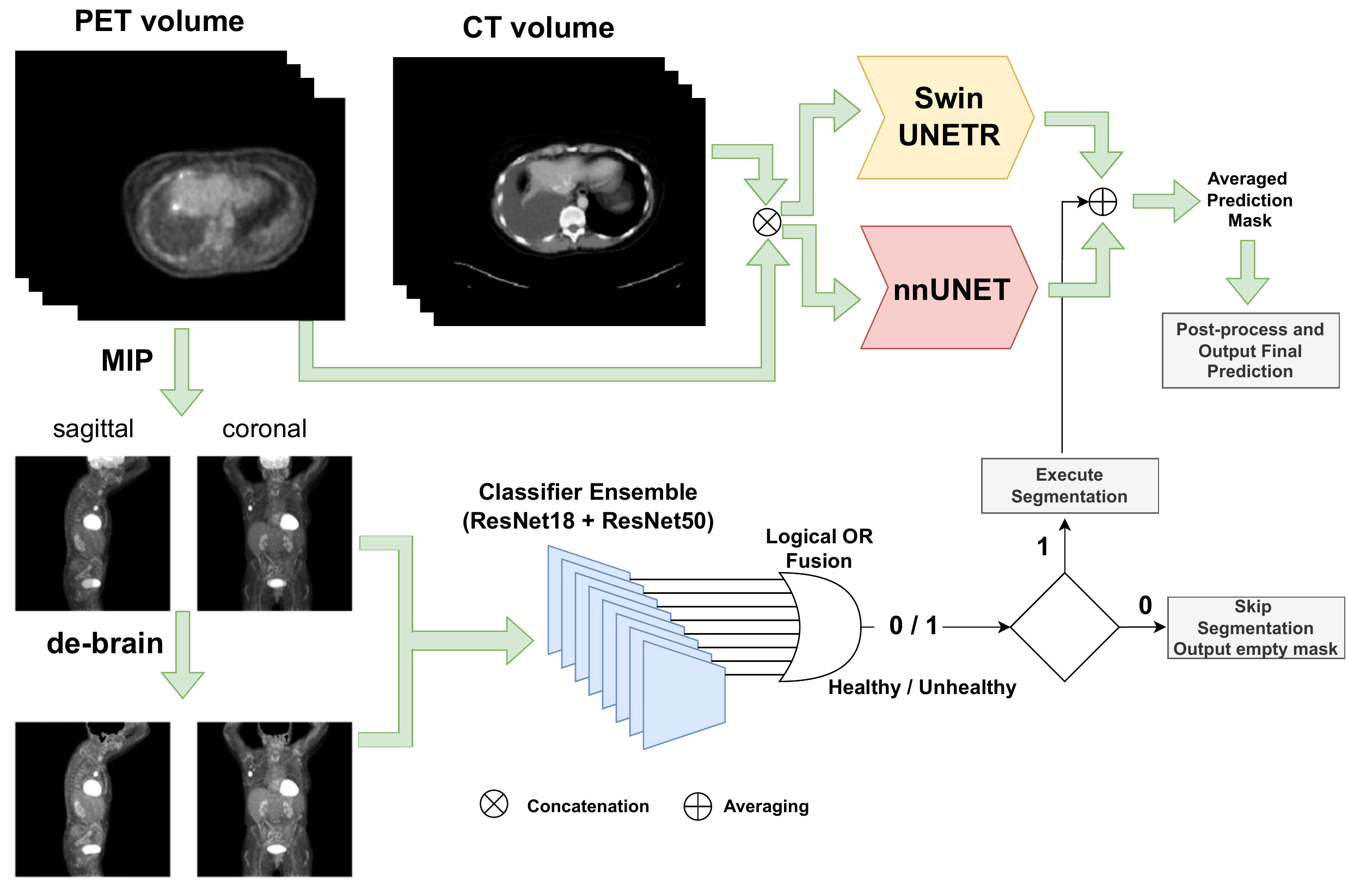}
    \caption{Our final submission pipeline, combining the Swin UNETR and the nn-Unet for segmentation with the ResNet18- and ResNet50-based classifiers.}
    \label{fig:pipeline}
\end{figure}
Our pipeline is illustrated in Figure~\ref{fig:pipeline}. The PET volume is used to compute sagittal and axial maximum intensity projections (MIPs) and their de-brained versions. These MIPs are fed to our ensemble of classifiers whose binary predictions are combined through a logical OR fusion. If all classifiers agree whether a patient has no pathological uptake, the segmentation process is skipped to avoid unnecessary computations and an empty prediction is used as an output. Otherwise, the PET and CT volumes are concatenated and used as an input to the nn-Unet and Swin UNETR models. Their predictions are fused by averaging and our post-processing (described in Section~\ref{sec:post-processing}) is applied to obtain the final prediction.
\subsection{Segmentation Models}
\label{sec:segmentation_models}

\subsubsection{nn-Unet -}

The nn-Unet \cite{nnUnet} is a powerful tool that can perform segmentation of various medical images. It is based on deep learning and automatically performs hyperparameter tuning, pre-processing, network architecture adjustment, training and post-processing without much manual intervention.






We explored two different architectures: 3D U-Net cascade and 3D full resolution U-Net.

\subsubsection{Swin UNETR -}
The recent success of vision transformers in image recognition \cite{dosovitskiy2020image,liu2021swin,liu2022swin} paved the way for attention-based architectures in the field of semantic segmentation \cite{cao2021swin}. With proposing the new architectures UNETR (UNEt TRansformers) and Swin UNETR, \cite{hatamizadeh2022unetr} and \cite{hatamizadeh2022swin} showed the competitiveness of vision transformers in 3D medical image segmentation with respect to the nn-Unet. Compared with fully convolutional neural networks, transformer based models shine at learning long-range information which is especially important for segmentation of tumors with variable size.\\

To provide further evidence of Swin UNETR's capabilities, we trained a model that processes FDG-PET/CT scans in a channel-wise manner. Our architecture processes a volume of size $(96, 96, 96)$ voxels and outputs two channels followed by a softmax. The feature size of the network was set to $48$. We optimized a Dice-Cross-Entropy loss (excluding the background) with an AdamW optimizer using a fixed learning rate of $0.0001$ and a weight decay of $0.00001$.

During training, we z-transformed the standardized uptake value (SUV) channel and performed percentile clipping and scaling to the CT channel such that its values are between zero and one. After foreground cropping, we randomly cropped a volume of size $(96, 96, 96)$ voxels, whose center is a lesion with probability $p_{lesion}=\frac{2}{3}$. Lastly, we randomly rotate the volume along each axis (with $p_{randrot}=0.2$). When performing inference, we used a sliding window inferer with an overlap of $0.25$ and gave equal weight to all predictions when blending the output of overlapping windows.\\

Besides segmenting lesions in scans that were diagnosed with cancer, the challenge focuses on the avoidance of false positive predictions in scans of healthy patients. We addressed the latter by attempting to classify whether lesions are existent in a scan. This approach is outlined below.

\subsection{Classification Models}
\label{Classification_Models}
To further improve the segmentation performance regarding healthy patients with empty ground truth masks, we incorporated binary tumor classification models into our segmentation pipeline. The core idea is that the classification task is a more simple objective than full pixelwise classification and a good classifier can de-noise an uncertain segmentation output. In essence, the classifier acts as a gating mechanism for the segmentation model. To this end, we investigated several backbones for classification and utilized the Maximum Intensity Projections (MIPs) of the PET volumes to train our classification models. We decided to use 2D MIPs instead of the whole volumes, as MIPs are typically used for tumor identification in standard clinical practice~\cite{georgi2022automatic} and our experiments showed that using MIPs can lead to sufficient accuracy ($\approx 95\%$) Code for the training can be found at: \url{https://github.com/Zrrr1997/autoPET-MirrorUnet}. 

\subsubsection{Objective of Classification}
As stated in Section \ref{sec:segmentation_models}, our objective in adding the classification models is to avoid the false negative cases, i.e. classify an unhealthy patient as healthy while still identifying obviously healthy patients and gating the segmentation model, i.e. multiplying its output by zero. Thus, instead of optimizing the accuracy, we opt for minimizing the false negatives with our classification model as a primary objective and minimizing the false positives as a secondary objective. This prioritization is necessary, so that our classification model does not discard correct segmentation predictions for unhealthy patients, which would dramatically decrease the Dice score. Instead, our goal is to build on top of the segmentation model and further enhance its performance. The necessity of a classification approach, which discerns between healthy and non-healthy patients, is amplified even further, given the fact that we use segmentation models trained only on non-healthy patients. 

We investigated three dimensions in varying our classifiers: their backbone architecture, the MIP-axis (sagittal and axial) which they are trained on, and whether the classifiers use MIPs with or without the brain removed. The brain is removed by applying a threshold filtering and removing the largest component with an uptake larger than the threshold hyperparameter. We show that forming an ensemble of classifiers trained with different configurations increases the generalizability of the models. 

\section{Results}

Given that the dataset is hidden during the model development we describe our experiments 
\subsection{Validation Strategy}
To make the best use of all the training data, we choose a 5-fold cross-validation (CV) scheme to split the data. Since some scans stem from the same patient  our split is grouped by the patient identifier to avoid information leakage. We utilized the provided metadata to make our splits stratified by sex and diagnosis to guarantee similar distributions across the folds.
\subsection{Segmentation Models}
\subsubsection{nn-Unet -} 
We trained the nn-Unet on the whole training data set as well as on the non-healthy scans only.  After training all folds, we found that the full resolution 3D U-net had the best overall Dice score when trained with only non-healthy patients, as seen in Table \ref{tab:nn-Unet}.
Since the 3D full resolution exhibits the best cross-validation with an average Dice score of $69.73\%$, see Table \ref{tab:nn-Unet}, we refer to it as \textit{the} nn-Unet in the subsequent sections. The false negative volume and false positive volume is listed in Table \ref{tab:nn-Unet_volumes}. \\
\begin{table}[]

\centering
\caption{Dice score of the nn-Unet for each network (3D Full Resolution and 3D Cascade) and each splitting (all patients and only non-healthy patients) trained and evaluated on non-healthy cases}
\begin{tabular}{|l|l|l|l|l|}
\hline
\textbf{Fold} & \textbf{3D FullRes} & \textbf{\begin{tabular}[c]{@{}l@{}}3D FullRes\\ (non-healthy)\end{tabular}} & \textbf{3D Cascade} & \textbf{\begin{tabular}[c]{@{}l@{}}3D Cascade \\ (non-healthy)\end{tabular}} \\ \hline
1 & 0.587 & 0.6876 & 0.6493 & 0.6357 \\ \hline
2 & 0.6467 & 0.6829 & 0.6636 & 0.6125 \\ \hline
3 & 0.7173 & 0.7133 & 0.5617 & 0.692 \\ \hline
4 & 0.7110 & 0.7275 & 0.6067 & 0.6675 \\ \hline
5 & 0.6211 & 0.6755 & 0.6474 & 0.5983 \\ \hline\hline
\textbf{CV} & \textbf{0.6566} & \textbf{0.6973} & \textbf{0.6258} & \textbf{0.6412} \\ \hline
\end{tabular}
\label{tab:nn-Unet}
\end{table}
\begin{table}[]
\centering
\caption{False Negative Volume (FNV) and False Positive Volume (FPV) of the 3D full resolution nn-Unet trained and evaluated on non-healthy cases}
\begin{tabular}{|c|c|c|}
\hline
\textbf{Fold} & \textbf{FNV} & \textbf{FPV} \\ \hline
1 & 4.3625 & 21.4024 \\ \hline
2 & 7.3647 & 17.6611 \\ \hline
3 & 28.3955 & 7.6103 \\ \hline
4 & 5.3868 & 13.9205 \\ \hline
5 & 6.994 & 11.1126 \\ \hline \hline
\textbf{CV} & \textbf{10.5007} & \textbf{14.3414} \\ \hline
\end{tabular}
\label{tab:nn-Unet_volumes}
\end{table}
\subsubsection{Swin UNETR -}We trained the Swin UNETR, for $150,000$ iterations, and evaluated the model each $2,000$ steps after iteration $70,000$. Inspired by the effect of distilling the signal by excluding healthy patients during training, our training data set was restricted to cases diagnosed with cancer. The cross-validation Dice scores are depicted in Table \ref{tab:Swin}. The Swin UNETR model is our second best model achieving a cross-validation Dice score of $66.75\%$.
\begin{table}[]
\centering
\caption{Dice Scores of Swin UNETR trained and evalutated on non-healthy cases}
\begin{tabular}{|c|c|}
\hline
\textbf{Fold} & \textbf{Dice} \\ \hline
1 & 0.6627 \\ \hline
2 & 0.6574 \\ \hline
3 & 0.6732 \\ \hline
4 & 0.6905 \\ \hline
5 & 0.6540 \\ \hline \hline
\textbf{CV} & \textbf{0.6675} \\ \hline
\end{tabular}
\label{tab:Swin}
\end{table}
\subsection{Late Fusion of nn-Unet and Swin UNETR}\label{sec:late_fusion}

Motivated by the hypothesis that both models learn complementary features, we applied late fusion by averaging the softmax probability outputs of both models. To evaluate the late fusion of the nn-Unet and the Swin UNETR, we computed the challenge's metrics on the training set (without healthy cases). The metrics are listed in Table \ref{tab:Ensemble}. The cross-validation metrics of the ensemble are better than our best single model (nn-Unet) in all considered metrics. Only the false negative volume of fold 1 is worse than its nn-Unet counterpart (cf. Table \ref{tab:nn-Unet_volumes}). 
\begin{table}[]
\centering
\caption{Dice Score, False Negative Volume (FNV), and False Positive Volume (FPV) of (nn-Unet + Swin UNETR) Late Fusion and the Cross Validation (CV) average}
\begin{tabular}{|c|c|c|c|}
\hline
\textbf{Fold} & \textbf{Dice} & \textbf{FNV} & \textbf{FPV} \\ \hline
1 & 0.7146 & 5.1083 & 8.5464 \\ \hline
2 & 0.6976 & 6.5240 & 7.8235 \\ \hline
3 & 0.7264 & 7.3805 & 4.8489 \\ \hline
4 & 0.7545 & 4.3221 & 10.7778 \\ \hline
5 & 0.7131 & 6.3486 & 7.6357 \\ \hline \hline
\textbf{CV} & \textbf{0.7212} & \textbf{5.9367} & \textbf{7.9265} \\ \hline
\end{tabular}
\label{tab:Ensemble}
\end{table}
\subsection{Classification Approach}
\subsubsection{Varying the Backbone Architecture -}
We trained and evaluated classification models based on three convolutional neural network (CNN) architectures: ResNet~\cite{he2016deep}, EfficientNet~\cite{tan2019efficientnet}, and CoAtNet~\cite{dai2021coatnet}. Our experiments showed that the ResNet model exhibits the most consistent and best performance on the classification task. The results regarding classification accuracy and FN/FP (false negative / false positive) ratio for fold 1 are illustrated in Table \ref{tab:classification_architectures} (results on other folds are left out for the sake of brevity). The results show that the classification models are able to reach an accuracy between $83\%-91\%$. However, only the ResNet50 and ResNet18 backbones achieve reasonable FN-rates, whereas EfficientNet and CoAtNet were not able to achieve fewer than four false positives. Hence, we opt to only use these two backbones for our late fusion ensemble experiments.

\subsubsection{Varying Brain/De-Brained Data -}
\begin{figure}[h]
    \centering
    \includegraphics[scale=0.8]{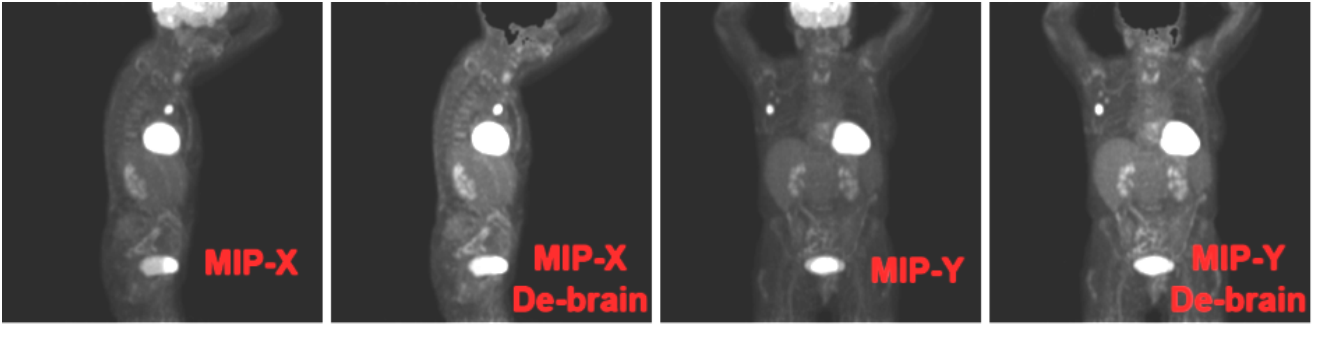}
    \caption{Example MIP images with and without the brain.}
    \label{fig:mip_examples}
\end{figure}

We trained additional classification models on MIPs with the brain removed as seen in Figure \ref{fig:mip_examples}. The MIPs show that tumor (positioned slightly above the heart) is clearly visible and removing the brain increases the global contrast in the image. We also speculate that removing the brain reduces the number of "distractions" with high metabolic uptake. The other two anatomical regions with high physiological uptake are the heart and the urinary bladder. However, removing them is dangerous, since there are patients whose tumors are in close proximity to these organs and a simple threshold filtering would remove the tumors as well. Table \ref{tab:classification_architectures} also shows the difference in the performance on the different MIP views (sagittal (X) and axial(Y)). The MIP-Y performance is consistently better than the MIP-X. 

\renewcommand{\arraystretch}{1.3}
\begin{table}[h]
\centering
\caption{Results for the architectures on fold $1$ with a decision threshold $\gamma=0.5$}
\scalebox{1.0}{
\begin{tabular}{|l|c|c|c|c|c|c|c|}
\hline
{} & \multicolumn{3}{c|}{\textbf{ResNet}} & \multicolumn{2}{c|}{\textbf{EfficientNet}} & \multicolumn{2}{c|}{\textbf{CoAtNet}} \\ \hline
\textbf{Backbone Version} & 18 & 50 & 101 & B0 & B4 & 3 & 4 \\ \hline
\textbf{Acc. MIP-X [\%]} & 83.66 & 87.62 & 87.13 & 83.66 & 89.60 & 87.13 & 86.63 \\ \hline
\textbf{Acc. MIP-Y [\%]} & 87.13 & 86.14 & 88.91 & 88.61 & 91.09 & 87.13 & 86.63 \\ \hline
\textbf{FN/FP MIP-X} & 14/33 & 16/23 & 8/32 & 16/22 & 4/32 & 6/29 & 10/33 \\ \hline
\textbf{FN/FP MIP-Y} & 10/24 & 2/65 & 1/49 & 4/28 & 10/19 & 11/12 & 6/32 \\ \hline
\end{tabular}}
\label{tab:classification_architectures}

\end{table}
We further analyze how the model's performance changes when using de-brained MIPs. The results are illustrated in Table \ref{tab:classification_debrain}. The results are similar to the ones with the brain, however we can observe that the models exhibit a \textit{different} distribution of FN/FP samples. This indicates that the models have learned diversified features and an ensemble of models trained on MIP data with brains and models trained on de-brained data would be beneficial.

\renewcommand{\arraystretch}{1.3}
\begin{table}[h]
\centering
\caption{Results for the architectures on fold 1 without the brain with $\gamma=0.5$}
\scalebox{1.0}{
\begin{tabular}{|l|c|c|c|c|c|c|c|}
\hline
{} & \multicolumn{3}{c|}{\textbf{ResNet}} & \multicolumn{2}{c|}{\textbf{EfficientNet}} & \multicolumn{2}{c|}{\textbf{CoAtNet}} \\ \hline
\textbf{Backbone Version} & 18 & 50 & 101 & B0 & B4 & 3 & 4 \\ \hline
\textbf{Acc. MIP-X [\%]} & 88.61 & 84.65 & 89.11 & 88.61 & 87.13 & 88.12 & 87.62 \\ \hline
\textbf{Acc. MIP-Y [\%]} & 87.62 & 87.62 &  84.16 & 88.12 & 90.10 & 89.60 & 88.61 \\ \hline
\textbf{FN/FP MIP-X} & 12/28 & 13/27 & 21/15 & 31/38 & 24/12 & 8/23 & 18/15 \\ \hline
\textbf{FN/FP MIP-Y} & 9/30 & 20/22 & 15/35 & 14/28 & 12/29 & 15/17 & 16/22 \\ \hline
\end{tabular}}
\label{tab:classification_debrain}

\end{table}

\subsubsection{Ensemble of Classifiers -}
We utilized the ResNet18 and ResNet50 backbones for our ensemble experiments as they showed the best FN/FP ratio, which is our primary objective. We ensemble based on three dimensions - trained on either MIP-X or MIP-Y data, trained with or without the brain, and using either ResNet18 or ResNet50 as a backbone, i.e. $2\times2\times2=8$ models for the ensemble. We employed decision-based late fusion, where we built the final prediction based on the logical \textbf{or} between the prediction of all classifiers. This means that all classifiers must agree on whether the patient is healthy in order to classify the patient as such. To further increase the robustness against false negatives, we used a conservative decision threshold for each classifier of $\gamma=0.3$. The final results can be seen in Table \ref{tab:final_classification_results}.

\renewcommand{\arraystretch}{1.3}
\begin{table}[h]
\centering
\caption{Results for the ensemble of all classifiers with $\gamma=0.3$}
\scalebox{1.0}{
\begin{tabular}{|l|c|c|c|c|c||c|}
\hline
{\textbf{Fold}} & 1 & 2 &  3 &  4 &  5 & \textbf{CV} \\ \hline
\textbf{Accuracy [\%]} & 69.8 & 70.9 & 77.9 & 73.8 & 79.4  & \textbf{74.3} \\ \hline
\textbf{FN/FP} & 1/60 & 2/58 & 4/39 & 3/50 & 2/41 & \textbf{2.4/49.6} \\ \hline
\end{tabular}}
\label{tab:final_classification_results}

\end{table}

\subsection{Post-Processing}
\label{sec:post-processing}
On the PET scans, sometimes image artifacts occur at the boundaries. Therefore, for the nn-Unet trained on all patients with tumors, we tested some post-processing steps. We evaluated the model by setting all predictions of the network at the boundaries to zero. To find out which boundary size works best, we set 0.5\%, 1\%, 1.5\%, 5\%, 10\%, 12\% of the scan on both ends of the $z$ axis to zero. We also evaluated to set a fixed number of slices to zero. The results only show marginal improvements. The results can be found in the Appendix A, Tables \ref{tab:low}, \ref{tab:up}, \ref{tab:Percentage_table}. 
Furthermore, we set all predictions with less that ten voxels in total to zero. Among all $900$ patients, this post-processing step only affected one patient with a tumor, whereas otherwise only false positive predictions in healthy patients were reduced. Nevertheless, the improvement was only marginal (cf. Appendix A, Table \ref{tab:FalsePositive}). 
\section{Conclusion}
In this contribution, we presented our approach for the AutoPET challenge. Our solution consists of two state-of-the-art segmentation models, namely nn-Unet and Swin UNETR, and a maximum intensity projections classifier that acts like a gating mechanism to the segmentation part of our pipeline. By performing a late fusion of both segmentation models we were able to boost the performance of the nn-Unet, a challenge-winning automated 3D segmentation framework. We suspect that this performance boost originated in the Swin UNETR's ability to learn long-range dependencies and the resulting complementary features. We hope to further investigate this assumption in the near future. Although we provided more evidence about the competitiveness of attention-based models in 3D semantic segmentation, our exploration of suitable hyperparameters for the Swin UNETR was limited. This will be explored in future work.
\section{Acknowledgment}
The present contribution is supported by the Cancer Research Center Cologne Essen (CCCE) and the Helmholtz Association under the joint research school “HIDSS4Health – Helmholtz Information and Data Science School for Health“.
%
%

%
%
%
%
\bibliographystyle{splncs04}
\bibliography{bib/refs}
\section*{Appendix}
\subsection*{Appendix A:}
The following tables show the results of the post-processing steps. 
\begin{table}[h]
\centering
\caption{Dices Scores if three, five or seven slices at the \textbf{lower boundary} of the $z$ axis are set to zero}
\begin{tabular}{|l|c|c|c|c|c|}
\hline
\textbf{Fold} & \textbf{1}  &  \textbf{2} & \textbf{3} &  \textbf{4} & \textbf{5}\\
\hline
original     & $0.687557$         &$\boldsymbol{0.682929}$ & $\boldsymbol{0.713271}$ & $0.727452$ & $0.675510$ \\
three slices & $0.687569$         &$0.682908$              & $0.713176$              & $0.727494$ & $0.675632$ \\
five slices & $0.687572$         &$0.682900$              & $0.713104$              & $0.727529$ & $0.675719$ \\
seven slices & $\boldsymbol{0.687573}$ &$0.682884$         & $0.713054$         & $\boldsymbol{0.727565}$ & $\boldsymbol{ 0.675781} $ \\
\hline
\end{tabular}
\label{tab:low}
\end{table}
\begin{table}[h]
\centering
\caption{Dices Scores if $3,5$ or $7$ slices at the \textbf{upper boundary} of the $z$ axis are set to zero}
\begin{tabular}{|l|c|c|c|c|c|}
\hline
\textbf{Fold} & \textbf{1}  & \textbf{2} & \textbf{3} & \textbf{4} & \textbf{5}\\
\hline
original     & $0.687557$              &$0.682929$              & $0.713271$              & $0.727452$ & $0.675510$ \\
three slices & $0.687562$              &$0.683581$              & $0.713285$              & $0.727453$ & $0.675513$ \\
five slices & $\boldsymbol{0.687565}$ &$0.683863$              & $\boldsymbol{0.713304}$ & $\boldsymbol{0.727454}$ & $0.675523$ \\
seven slices & $\boldsymbol{0.687565}$ &$\boldsymbol{0.68388}$  & $0.713272$              & $\boldsymbol{0.727454}$ & $\boldsymbol{ 0.675542} $ \\
\hline
\end{tabular}
\label{tab:up}
\end{table}
\begin{table}[h]
\centering
\caption{Dices Scores if $0.5\%,1\%,1.5\%,2\%,5\%,10\%,12\%$ or $15\%$ of the image size was set to zero at both boundaries of the $z$ axis}
\begin{tabular}{|l|c|c|c|c|c|}
\hline
\textbf{Fold}  & \textbf{1}  & \textbf{2} & \textbf{3} & \textbf{4} & \textbf{5}\\
\hline
original   & $0.687557$               & $0.682929$            & $\boldsymbol{0.713271}$ & $0.727452$               & $0.675510$ \\
0.5\%  & $0.687562$               & $0.683303$            & $0.713264$              & $0.727461$               & $0.675543$ \\
1  \%  & $0.687572$               & $0.683800$              & $0.713190$              & $0.727481$               & $0.675634$\\
1.5\%  & $0.687581$               & $0.683869$            & $0.713156$              & $\boldsymbol{0.727514}$  & $0.675685$ \\
2  \%  & $0.687582$               & $0.684290$            & $0.713125$              & $0.727531$               & $0.675794$ \\
5  \%  & $0.687556$               & $\boldsymbol{0.684400}$ & $0.712817$              & $0.720657$               & $0.67.6615$ \\
10 \%  & $0.688816$               & $0.670933$            & $0.711193$              & $0.71.9549$               & $\boldsymbol{0.678074}$ \\
12 \%  & $\boldsymbol{0.688831}$  & $0.661311$            & $0.711827$              & $0.718250$                & $0.677589$ \\
15 \%  & $0.681990$                & $0.655481$            & $0.704031$              & $0.714196$               & $0.667331$ \\
\hline
\end{tabular}
\label{tab:Percentage_table}
\end{table}
\begin{table}[h]
\centering
\caption{False Positive Volume after predictions less than or equal to ten set to zero, including healthy and non-healthy patients in the validation sets}
\begin{tabular}{|l|c|c|c|c|c|}
\hline
\textbf{Fold} & \textbf{1}  & \textbf{2} & \textbf{3} & \textbf{4} & \textbf{5}\\
\hline
original & $22.3454$ &$21.6614$ & $8.2080$ & $15.3581$ & $29.1173$ \\
post-processed  & $\textbf{22.3448}$ &$\textbf{21.6601}$ & $\textbf{8.2047}$ & $\textbf{15.3556}$ & $\textbf{29.1160}$ \\
\hline
\end{tabular}
\label{tab:FalsePositive}
\end{table}
\end{document}